\documentclass[12pt,a4paper]{article}

\usepackage[utf8]{inputenc}

\usepackage{amsmath,amssymb,amsthm}
\usepackage{mathtools}

\usepackage{geometry}
\usepackage{graphicx}

\usepackage{tikz}
\usetikzlibrary{arrows.meta,positioning,calc}

\usepackage{caption}

\usepackage[numbers]{natbib}

\usepackage{hyperref}

\usepackage{setspace}
\hypersetup{
    colorlinks=true,
    linkcolor=blue,
    citecolor=blue,
    urlcolor=blue
}

\title{
\textbf{
Why Efficient Reforms Fail:\\
Endogenous Game Transformation under Status Quo Bias and Social Preferences
}
}

\author{
Madjid Eshaghi Gordji\thanks{
Corresponding author:
\href{mailto:meshaghi@semnan.ac.ir}{meshaghi@semnan.ac.ir}
}
\and
Mohammadali Berahman
\and
Hasti Eshaghi
}

\date{}

\newcommand{\SemnanAffiliation}{
Faculty of Mathematics, Statistics and Computer Science,\\
Semnan University, Semnan, Iran
}

\newcommand{\TehranAffiliation}{
Faculty of Mathematics,\\
University of Tehran, Tehran, Iran
}

\newtheorem{theorem}{Theorem}[section]
\newtheorem{proposition}[theorem]{Proposition}

\theoremstyle{definition}

\begin{document}

\maketitle

\begin{center}

\textbf{Madjid Eshaghi Gordji}\\
\SemnanAffiliation\\
\texttt{meshaghi@semnan.ac.ir}

\vspace{0.5cm}

\textbf{Mohammadali Berahman}\\
\SemnanAffiliation

\vspace{0.5cm}

\textbf{Hasti Eshaghi}\\
\TehranAffiliation

\end{center}


\begin{abstract}
Why do societies remain stuck in inferior institutions even when superior alternatives are widely recognized? This paper develops a model in which agents choose not only actions within a game but also transformations of the game itself. Transformations may be \emph{soft}, changing payoffs through taxes or subsidies, or \emph{hard}, changing feasibility through deletion or replacement of actions. Within a coordination model with status-quo bias (switching cost) and boundedly rational play (logit quantal response), we show that these interventions are qualitatively different: finite taxes shift behavior continuously but cannot eliminate residual use of the inherited action, whereas deletion bypasses inertia by removing the action from the feasible set. We further characterize how antagonistic social preferences at the meta level can block reforms that are individually beneficial for every player. The framework provides a formal rationale for why hard feasibility restrictions often dominate soft price incentives under inertia, with direct applications to climate transition (carbon tax vs. fossil-fuel phase-out) and platform regulation (fines vs. deletion of addictive features).
\end{abstract}

\noindent\textbf{Keywords:} Bounded rationality, quantal response, switching costs, status quo bias, institutional change, action deletion.

\section{Introduction}

The persistence of inferior institutions, technologies, and social norms is visible in settings ranging from fossil-fuel dependence and unsafe platform design to arms races in AI deployment. Standard explanations include coordination failure \citep{Cooper1999}, path dependence \citep{Arthur1989,North1990}, and strategic uncertainty \citep{MorrisShin1998}. Behavioral economists have further documented a robust \emph{status quo bias}: individuals disproportionately stick with inherited choices even when switching would improve their welfare \citep{SamuelsonZeckhauser1988,KahnemanKnetschThaler1991}. Meanwhile, the bounded rationality literature emphasizes that real decision makers do not fully optimize but respond noisily to incentives \citep{Simon1955,Selten1998,Camerer2003}. The quantal response equilibrium (QRE) \citep{McKelveyPalfrey1995} provides a smooth, tractable framework for such noisy behavior and has been widely applied in experimental and theoretical work \citep{GoereeHoltPalfrey2016}.

This paper addresses a complementary and largely overlooked dimension: players can choose \emph{which game to play}. Institutions, policies, and technologies are not fixed; they are themselves objects of strategic choice. A policymaker can impose a tax (price-only reform) or delete an action entirely (ban). Players can propose such transformations, vote on them, or lobby for them. The question then becomes: under bounded rationality and status-quo inertia, when do price interventions succeed and when do they fail? And how do social preferences (friendship, enmity) at the meta level affect reform outcomes? This connects to a growing literature on endogenous institutions \citep{AcemogluRobinson2006,Greif2006} and behavioral mechanism design \citep{Sutter2009,Kamenica2012}.
We develop a framework that integrates three elements: bounded rationality via QRE with a switching cost (status-quo bias) \citep{DubinMcFadden1984,FudenbergLevine2006}, endogenous game transformations (price-only, deletion, addition, replacement) \citep{Hurwicz2008,MaskinTirole2001}, and social preferences at the meta level, extending the hyper-rationality concept of \citet{AskariEshaghiPark2019} from ordinary actions to rule-changing acts. We define two equilibrium concepts: \emph{Meta-Nash equilibrium} (Nash equilibrium of the induced meta-game) and \emph{Hyper-Meta-Nash equilibrium} (same but with hyper-rational social utilities). The latter relates to literature on other-regarding preferences \citep{FehrSchmidt1999,BoltonOckenfels2000} and to political economy of reform \citep{FernandezRodrik1991,AlesinaDrazen1991}.
Our main results are as follows. First, we prove comparative statics: in a binary coordination game with QRE-SB, the probability of choosing the status quo increases with switching cost and decreases with the tax on the status quo action. Second, we show that no finite tax can replicate the effect of deletion: under any full-support decision rule (logit being a special case), price-only interventions keep every feasible action with positive probability, while deletion forces zero probability. Third, we establish that arbitrarily high taxes can approximate deletion but never achieve it exactly. Fourth, we demonstrate that a single sufficiently spiteful player can block a Pareto-improving reform under unanimity, whereas majority voting may overcome such blocking. Finally, we provide a welfare-based cost threshold for when deletion dominates taxation. These results are applied to two pressing policy problems: climate transition (carbon tax vs. fossil-fuel phase-out) and platform regulation (fines vs. deletion of addictive features), both of which have received recent attention \citep{Nordhaus2015,AllcottGentzkow2017,Alter2017}.
The remainder of the paper is organized as follows. Section 2 lays out the formal model, including QRE-SB, game transformations, and the induced meta-game. Section 3 defines Meta-Nash and Hyper-Meta-Nash equilibria. Section 4 presents the binary coordination model and derives the main theorems. Section 5 provides numerical illustrations. Section 6 discusses the broad applicability of the framework and then zooms in on two detailed applications. Section 7 concludes. All proofs are collected in the Appendix.

\section{Model Framework}

\subsection{Basic Game and Bounded Rationality}

A finite normal-form game is \(G = (N, \{A_i\}_{i\in N}, \{u_i\}_{i\in N})\) with \(N=\{1,\dots,m\}\). We introduce a \textbf{quantal response equilibrium with status-quo bias (QRE-SB)}. Let \(a_i^0\) be player \(i\)'s default action (the status quo). A switching cost \(\kappa_i \ge 0\) is incurred whenever player \(i\) chooses an action different from \(a_i^0\). Given beliefs about others' strategies \(\sigma_{-i}\), the effective payoff is
\[
\tilde u_i(a_i, \sigma_{-i}) = \mathbb{E}_{a_{-i}\sim \sigma_{-i}}[u_i(a_i,a_{-i})] - \kappa_i \mathbf{1}_{\{a_i\neq a_i^0\}}.
\]

The logit choice probability (precision \(\beta_i\ge 0\)) is
\[
\sigma_i(a_i) = \frac{\exp\bigl(\beta_i \tilde u_i(a_i,\sigma_{-i})\bigr)}{\sum_{a_i'\in A_i}\exp\bigl(\beta_i \tilde u_i(a_i',\sigma_{-i})\bigr)}.
\]

A QRE-SB is a fixed point of this system. For symmetric games with symmetric default, we look for symmetric equilibria. The parameter \(\beta_i\) governs the sensitivity to payoff differences: as \(\beta_i\to 0\), choice becomes uniform; as \(\beta_i\to\infty\), it converges to deterministic best response with switching cost. Finite \(\beta_i\) captures bounded rationality, while \(\kappa_i\) captures inertia or status-quo attachment.
\subsection{Game Transformations}

Let \(\mathcal{G}\) be the set of all games with the same player set. A \textbf{transformation} is an operator \(T:\mathcal{G}\to\mathcal{G}\). We focus on four types, which cover a broad and economically important subset of reforms that alter payoffs or feasible action sets: price-only (\(T^P\)), deletion (\(T^D\)), addition (\(T^A\)), and replacement (\(T^R\)). The transformation rule \(\Phi\) introduced below is intentionally reduced form; it can represent unanimity requirements, majority voting, bargaining protocols, or centralized regulatory choice. The economic content lies in the mapping from meta-actions to transformed games.
\subsection{The Induced Meta-Game}

Let \(G_0\) be the initial game. Each player \(i\) chooses a \textbf{meta-action} \(x_i \in X_i\) (finite set). The environment chooses \(e \in E\) (finite). A transformation rule \(\Phi: X_1\times\cdots\times X_m\times E \to \mathcal{T}\) maps every profile \(y=(x_1,\dots,x_m,e)\) to a concrete transformation. The transformed game is \(G(y)=\Phi(y)(G_0)\). Let \(\Gamma\) be a \textbf{within-game equilibrium selection rule} that assigns to each game a unique outcome (e.g., the unique QRE-SB). Denote by \(\Pi_i(\Gamma(G))\) the expected payoff of player \(i\) in that outcome. The \textbf{meta-payoff} of player \(i\) is \(V_i(y) = \Pi_i(\Gamma(G(y))) - C_i(y)\), where \(C_i(y)\ge 0\) captures the cost of proposing or implementing the transformation (e.g., lobbying, enforcement). A within-game selection rule \(\Gamma\) is admissible if it returns a unique outcome and the induced payoff vector is well-defined.
\section{Equilibrium Concepts at the Meta Level}

\subsection{Meta-Nash Equilibrium}

A profile \(y^*=(x_1^*,\dots,x_m^*,e^*)\) is a \textbf{Meta-Nash equilibrium} if for every player \(i\) and every \(x_i\in X_i\), \(V_i(x_i^*,x_{-i}^*,e^*) \ge V_i(x_i,x_{-i}^*,e^*)\), and if the environment is strategic with payoff \(V_0\), then for every \(e\in E\), \(V_0(x^*,e^*) \ge V_0(x^*,e)\). This is formally a Nash equilibrium of an induced strategic-form game, but economically novel because players optimize over rule-changing acts, not merely over within-game actions.

\subsection{Hyper-Meta-Nash Equilibrium}

Following \citet{AskariEshaghiPark2019}, we incorporate social preferences. Each player \(i\) has weights \(\omega_{ij}\in\mathbb{R}\) for \(j\neq i\) (positive = friendship, negative = enmity). The \textbf{hyper-meta-payoff} is
\[
H_i(y) = V_i(y) + \sum_{j\neq i} \omega_{ij} V_j(y).
\]

A profile \(y^*\) is a \textbf{Hyper-Meta-Nash equilibrium} if for every player \(i\) and every \(x_i\in X_i\),
\[
H_i(x_i^*,x_{-i}^*,e^*) \ge H_i(x_i,x_{-i}^*,e^*),
\]
(with analogous condition for the environment if strategic).
\subsection{Existence}

If \(X_i\) and \(E\) are finite and \(\Gamma\) is admissible, then a mixed-strategy Meta-Nash equilibrium and a mixed-strategy Hyper-Meta-Nash equilibrium exist, because the meta-game is a finite normal-form game and Nash's theorem applies.
\section{Application: Binary Coordination with Status-Quo Bias}

\subsection{The Coordination Game}

Consider a symmetric two-player binary coordination game with payoff matrix:
\[
\begin{array}{c|cc}
 & X & Y \\ \hline
X & a,\;a & c,\;d \\
Y & d,\;c & b,\;b
\end{array}
\]

Assumptions: \(a>c,\ b>d,\ b>a\) ( \((Y,Y)\) Pareto-superior). Status quo default is \((X,X)\) with switching cost \(\kappa\ge0\) for deviating from \(X\). Let \(p\) be the probability that a player chooses \(X\) in a symmetric mixed strategy. Expected payoffs are
\[
U_X(p)=pa+(1-p)c
\]
and
\[
U_Y(p)=pd+(1-p)b-\kappa.
\]

The difference \(U_Y(p)-U_X(p)\) simplifies to
\[
\Delta(p) = \alpha - \kappa - p(\alpha+\gamma),
\]
where \(\alpha=b-c>0\) and \(\gamma=a-d>0\). The symmetric QRE-SB condition is
\[
p = \frac{1}{1+\exp(\beta \Delta(p))}.
\]

Under the contraction condition \(\beta(\alpha+\gamma)<4\), the map is a contraction on \([0,1]\) and the equilibrium is unique.
A tax \(t\ge0\) on action \(X\) (price-only) reduces the payoff from \(X\) by \(t\), so
\[
\Delta_t(p)=\alpha-\kappa-p(\alpha+\gamma)+t.
\]

The critical tax that makes \(p=1/2\) is
\[
\bar t = \kappa - \frac{\alpha-\gamma}{2}.
\]

If \(t<\bar t\) then \(p>1/2\); if \(t>\bar t\) then \(p<1/2\). Deleting action \(X\) removes it from the feasible set, forcing \(p=0\).
\subsection{Main Theorems}

\begin{theorem}[Comparative Statics of Status-Quo Persistence]\label{th:comp}
Under the contraction condition, the unique equilibrium probability \(p^*\) satisfies:
\[
\frac{\partial p^*}{\partial \kappa}>0,\quad
\frac{\partial p^*}{\partial t}<0,\quad
\frac{\partial p^*}{\partial \alpha}<0,\quad
\frac{\partial p^*}{\partial \gamma}>0.
\]
\end{theorem}

The proof is given in the Appendix; it uses the implicit function theorem and the fact that the logistic derivative is bounded.

\begin{theorem}[No Finite Tax Replicates Deletion]\label{th:nodelete}
Under logit QRE-SB with finite precision \(\beta<\infty\), if action \(X\) remains feasible then for every finite tax \(t\) we have \(p_t>0\). Deleting \(X\) yields \(p_D=0\). Hence no finite price-only intervention can replicate the behavioral effect of deletion.
\end{theorem}

The proof follows directly from the strict positivity of logit probabilities for feasible actions.

\begin{theorem}[Approximation without Replication]\label{th:approx}
Under logit QRE-SB with finite \(\beta<\infty\), \(\lim_{t\to\infty} p_t = 0\), but for every finite \(t\), \(p_t>0\). Hence arbitrarily high taxes can approximate the effect of deletion but never achieve it exactly.
\end{theorem}

This is proved by noting that the exponent in the logit map tends to \(+\infty\) as \(t\to\infty\).

\begin{theorem}[A Single Antagonistic Player Can Block Efficient Reform]\label{th:block}
Consider an \(n\)-player unanimity rule under which a reform is implemented iff every player chooses Approve. Let reform payoffs be \(U^D=(U_1^D,\dots,U_n^D)\) and status quo \(U^S=(U_1^S,\dots,U_n^S)\) with \(U_i^D>U_i^S\) for all \(i\). If there exists a player \(i\) such that
\[
\sum_{j\neq i}\omega_{ij}(U_j^D-U_j^S)<-(U_i^D-U_i^S),
\]
then unanimous approval cannot be a Hyper-Meta-Nash equilibrium.
\end{theorem}

The proof calculates the change in hyper-meta-payoff for player \(i\) and shows it is negative under the stated condition.
\begin{theorem}[Efficient Reform as a Meta-Nash Equilibrium]\label{th:metareform}
Under unanimity, suppose for every player \(i\),
\[
U_i^D-U_i^S > C_i(\text{Approve},\ldots,\text{Approve})-C_i(\text{Reject},\ldots,\text{Reject}).
\]

Then the profile in which every player chooses Approve is a Meta-Nash equilibrium.
\end{theorem}

This follows because unilateral deviation reduces net payoff.

\begin{proposition}[Qualitative Non-Equivalence of Soft and Hard Interventions]\label{prop:noneq}
Under finite-precision logit QRE-SB, any price-only intervention preserves full support over feasible actions, whereas any deletion intervention can assign zero probability to the deleted action. Thus price-only and feasibility-changing interventions are qualitatively non-equivalent.
\end{proposition}
\section{Numerical Illustrations}

Set \(a=6,\ b=7,\ c=1,\ d=2\), so \(\alpha=6,\ \gamma=4\). Choose \(\kappa=1.5\) giving \(\bar t=0.5\). With \(\beta=1\):

\[
\begin{array}{c|c|c}
t & p_t & SW(p_t) \\ \hline
0 & 0.78 & 6.34 \\
0.5 & 0.50 & 6.25 \\
1.0 & 0.22 & 6.20 \\
\infty\ (\text{deletion}) & 0 & 7.00
\end{array}
\]

Figure 1 shows the equilibrium probability as a function of the tax; the vertical dashed line marks \(\bar t=0.5\); deletion corresponds to the limit \(p\to0\).

\begin{figure}[h]
\centering
\begin{tikzpicture}[scale=1.2]

\draw[->] (0,0) -- (5,0) node[right] {$t$};
\draw[->] (0,0) -- (0,4) node[above] {$p_t$};

\draw[thick] 
(0,0.78) -- (0.5,0.5) -- (1.5,0.22) -- (4,0.05);

\draw[dashed] 
(0.5,0) -- (0.5,0.5) -- (0,0.5);

\node at (0.5,-0.1) {$\bar t$};
\node at (0.8,0.6) {$p=1/2$};

\end{tikzpicture}

\caption{Equilibrium probability of status quo as a function of tax.}
\label{fig:threshold}

\end{figure}
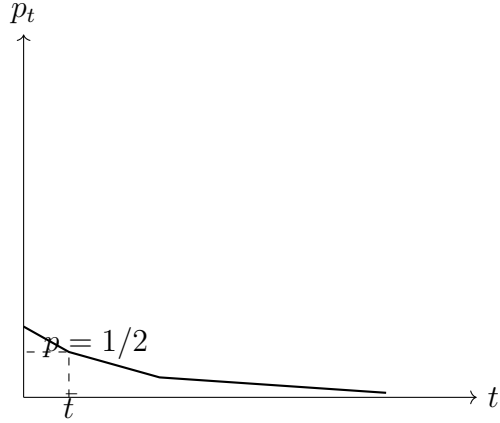

\section{Applications}

The framework developed above is remarkably versatile. It applies to any strategic setting where (i) there is an inherited status quo, (ii) agents are boundedly rational and averse to switching, (iii) they can propose changes to the rules or available actions, and (iv) they may care about each other's payoffs. Examples include technology adoption, international environmental agreements, corporate governance, platform design, public health interventions, and even military alliances. In each case, the core distinction between soft incentives (taxes, subsidies, fines) and hard constraints (deletion of actions, bans, mandated standards) is crucial. Our theorems show that soft incentives only shift behavior continuously, while hard constraints can eliminate the undesirable action entirely. Moreover, when social preferences are present (friendship or enmity), even Pareto‑improving reforms can be blocked under unanimity (Theorem 4), but majority voting may rescue them (Corollary to Theorem 4). Below we elaborate on two concrete, high‑impact applications.
\subsection{Climate Transition: Carbon Tax versus Fossil-Fuel Phase-Out}

The model can be directly mapped to the problem of decarbonisation. Let \(X\) represent the continued use of a legacy fossil-fuel technology (e.g., internal combustion engine vehicles, coal-fired power plants) and \(Y\) a cleaner alternative (electric vehicles, renewables). Adoption of \(Y\) exhibits strategic complementarities: more users lead to better charging infrastructure, lower costs, and stronger political support. The status quo is locked in by switching costs \(\kappa\): sunk capital, retooling expenses, training requirements, and contractual obligations. A carbon tax is a price-only intervention: it raises the immediate cost of using \(X\) but leaves the option \(X\) available. A statutory phase-out (e.g., the EU's 2035 ban on new petrol and diesel cars) is a deletion transformation: it removes the fossil-fuel option from the feasible set. Our Proposition~\ref{prop:noneq} (Qualitative Non-Equivalence) implies that under any finite carbon tax, the probability of using fossil fuels remains strictly positive, whereas a phase-out forces this probability to zero. Theorem~\ref{th:approx} (Approximation without Replication) shows that even an arbitrarily high carbon tax can only approach zero, never hit it exactly. Thus the model explains why carbon pricing alone is often insufficient to achieve complete decarbonisation, and why complementary bans or technology mandates are increasingly advocated. Furthermore, Theorem~\ref{th:block} (Antagonistic Player Blocking Reform) can be interpreted as a political-economy channel: if a fossil-fuel incumbent (e.g., an oil company or a coal region) has sufficiently negative weight on the payoffs of green adopters (i.e., spite), it can veto a unanimous reform. This suggests that moving from unanimity to majority voting (or side payments) may be necessary to overcome such blocking.
\subsection{Platform Regulation: Fines versus Deletion of Addictive Features}

Our second application concerns digital platforms. Let \(X\) be an engagement-maximising but potentially harmful design feature, such as infinite scroll, autoplay, targeted advertising to minors, or "like" counts. Let \(Y\) be a safer design (e.g., friction-based interfaces, age-verified feeds). A platform that unilaterally switches to \(Y\) risks losing user engagement and advertising revenue relative to competitors that retain \(X\); thus there is a coordination problem. The status quo is sustained by switching costs \(\kappa\): development costs, fear of user churn, and internal inertia. A financial penalty per unit of harmful engagement (e.g., a fine under the EU Digital Services Act) is a price-only intervention: it raises the cost of using \(X\) but keeps it feasible. Deletion of the feature (e.g., a legal ban on infinite scroll for minors) is a hard intervention that removes \(X\) from the action set.
Our Theorem~\ref{th:nodelete} (No Finite Tax Replicates Deletion) implies that no finite fine can drive the probability of using the harmful feature exactly to zero; only deletion can. Theorem~\ref{th:approx} shows that an extremely large fine could approximate deletion, but in practice such fines may be infeasible or politically unacceptable. Therefore the model provides a formal justification for design-based regulation that directly prohibits certain features rather than merely taxing them. Moreover, Theorem~\ref{th:block} (Antagonistic Player Blocking Reform) applies to platform governance: if a single member of a board or a dominant shareholder has a strong enmity toward safer design (e.g., because they benefit from addiction-driven advertising revenue), they can block a unanimous move to delete the feature. This highlights the importance of voting rules and the potential need for regulatory mandates that override internal unanimity requirements.

In both applications, the key insight is the same: under bounded rationality and status-quo bias, feasibility-changing interventions (deletion) are qualitatively more powerful than price-only interventions. The model thus gives a rigorous foundation for the policy principle that ``sometimes you must ban, not just tax.'' It also shows how social preferences can create additional barriers to reform, and how the design of voting rules (unanimity vs. majority) matters crucially.

\section{Conclusion}

We have introduced a framework in which agents strategically choose game transformations under bounded rationality and social preferences. In a binary coordination model with status-quo bias, we proved that price-only interventions can only shift behavior continuously, whereas feasibility-changing interventions can eliminate the inherited action entirely. We further showed that a single spiteful player can block a Pareto-improving reform, and that efficient reform is achievable as a Meta-Nash equilibrium when implementation costs are low. The analysis provides a formal foundation for policy debates on climate transition and platform regulation. Future work will extend the model to dynamic learning, network interactions, and empirical calibration.

\appendix

\section{Proof of Theorem \ref{th:comp}}

Define \(F(p;t,\kappa,\alpha,\gamma)=p-\frac{1}{1+\exp\{\beta[\alpha-\kappa-p(\alpha+\gamma)+t]\}}\). Let \(L(p)=\frac{1}{1+\exp\{\beta[\alpha-\kappa-p(\alpha+\gamma)+t]\}}\). Then \(F(p)=p-L(p)\). Differentiating with respect to \(p\): \(\frac{\partial L}{\partial p}=\beta(\alpha+\gamma)L(1-L)\), so \(F_p=1-\beta(\alpha+\gamma)L(1-L)\). Since \(L(1-L)\le 1/4\), the condition \(\beta(\alpha+\gamma)<4\) implies \(F_p>0\). Hence the implicit function theorem applies. Compute \(\frac{\partial L}{\partial \kappa}=\beta L(1-L)\), \(\frac{\partial L}{\partial t}=-\beta L(1-L)\), \(\frac{\partial L}{\partial \alpha}=-\beta(1-p)L(1-L)\), \(\frac{\partial L}{\partial \gamma}=\beta pL(1-L)\). Then \(F_\kappa=-\beta L(1-L)<0\), \(F_t=\beta L(1-L)>0\), \(F_\alpha=\beta(1-p)L(1-L)>0\), \(F_\gamma=-\beta pL(1-L)<0\). Using \(\partial p^*/\partial \theta = -F_\theta/F_p\) gives the signs.


\section*{Conflict of Interest}

The authors declare that they have no conflict of interest.
\section*{Funding}

The authors received no financial support for the research, authorship, or publication of this article.

\section*{Use of AI Assistance}

The authors used AI-based language assistance tools solely for improving grammar, readability, and formatting of the manuscript. All scientific content, mathematical results, proofs, and interpretations were developed and verified by the authors.

\bibliographystyle{unsrtnat}
\bibliography{references}

\end{document}